\documentclass[aps,prd,onecolumn,superscriptaddress,nofootinbib,showpacs]{revtex4}

\usepackage{graphicx}
\usepackage{slashed}
\usepackage{amsmath,bbm,latexsym,amssymb}
\usepackage{epsfig}
\usepackage{epstopdf}

\newcommand{\be}{\begin{equation}}
\newcommand{\ee}{\end{equation}}
\newcommand{\ba}{\begin{eqnarray}}
\newcommand{\ea}{\end{eqnarray}}

\begin{document}
\begin{flushright}
CFTP/21-004\\[-1mm]
\end{flushright}
\vspace*{1cm}

\title{Symmetry and decoupling in multi Higgs models}

\author{Sergio Carrolo}\thanks{E-mail: sergio.carrolo@tecnico.ulisboa.pt}
\affiliation{CFTP, Departamento de F\'{\i}sica,
Instituto Superior T\'{e}cnico, Universidade de Lisboa,
Avenida Rovisco Pais 1, 1049 Lisboa, Portugal}
\author{Jorge C.\ Rom\~{a}o}\thanks{E-mail: jorge.romao@tecnico.ulisboa.pt}
\affiliation{CFTP, Departamento de F\'{\i}sica,
Instituto Superior T\'{e}cnico, Universidade de Lisboa,
Avenida Rovisco Pais 1, 1049 Lisboa, Portugal}
\author{Jo\~{a}o P.\ Silva}\thanks{E-mail: jpsilva@cftp.ist.utl.pt}
\affiliation{CFTP, Departamento de F\'{\i}sica,
Instituto Superior T\'{e}cnico, Universidade de Lisboa,
Avenida Rovisco Pais 1, 1049 Lisboa, Portugal}
\author{Francisco Vaz\~{a}o}\thanks{E-mail: francisco.vazao@tecnico.ulisboa.pt}
\affiliation{CFTP, Departamento de F\'{\i}sica,
Instituto Superior T\'{e}cnico, Universidade de Lisboa,
Avenida Rovisco Pais 1, 1049 Lisboa, Portugal}

\date{\today}

\begin{abstract}
We study the scalar sector of the most general multi Higgs doublet model
satisfying explicitly an \textit{exact} symmetry. We prove that
such a model will exhibit decoupling if and only if the vacuum
also satisfies the same symmetry.
This general property is also shown independently and explicitly for
three Higgs doublet models, by considering in detail
all symmetry-constrained models and their possible vacua.
We also discuss some specific characteristics of different decoupling
patterns.
\end{abstract}


\maketitle

\section{\label{sec:intro}Introduction}

There is great interest in models with extra scalars, for they can correct many of
the shortcomings of the Standard Model (SM), such as the need for extra
sources of CP violation to drive baryogenesis, the need for Dark Matter,
or even, for example through a hierarchy of vacuum expectation values (vev),
explain the smallness of neutrino masses.
An independent source of interest lies in the fact that the ATLAS and CMS
collaborations
\cite{Aad:2012tfa,Chatrchyan:2012xdj}
have found a fundamental scalar (the 125GeV Higgs boson $h_{125}$),
prompting the obvious question:
how many fundamental scalars are there in Nature?
Will it happen as in the fermion sector, where there are multiple families?
As a result, many articles focus on N Higgs doublet models (NHDM) --
for reviews, see, for example \cite{hhg,Branco:2011iw,Ivanov:2017dad},
and references therein.

But,
multi scalar models are already constrained by
data from LHC. In particular,
the couplings of the $h_{125}$ to gauge bosons and the heaviest
charged fermions are known to coincide with couplings expected in the SM,
with errors of order 20\% or better
\cite{Aad:2019mbh,Sirunyan:2018koj,CMS:2020gsy,ATLAS:2020qdt}.
This feature is easy to explain in models which have a so-called
\textit{decoupling limit} \cite{Gunion:2002zf}.
In that limit, the extra scalar fields have large masses and
what is left at low energy is a state whose properties approach
naturally those of the SM Higgs boson.
The most general NHDM does have a decoupling limit.
However, it has too many parameters and may suffer from
large flavour changing neutral scalar couplings,
which are very constrained by flavour physics experiments.
So, it has become commonplace to add specific
symmetries to the NHDM.

Nevertheless,
it has been appreciated for a while that many such
symmetry-constrained models \textit{cannot} accommodate a decoupling limit;
see, for example,
\cite{Gunion:2002zf,Bhattacharyya:2014oka,Nebot:2019lzf,Nierste:2019fbx}.
In this article,
we use a very general method to show that
for \textit{any} NHDM with an \textit{exact} symmetry,
there will be a decoupling limit if and only if the vacuum also
respects that symmetry.
In contrast,
reference \cite{Faro:2020qyp} considered
partial results applicable to NHDM with
\textit{soft symmetry breaking}.\footnote{It is interesting to note that
that also bounded from below conditions deduced for the case with
an exact symmetry can be invalidated by the
introduction of soft symmetry breaking terms \cite{Ivanov:2020jra}.}

We present the notation in section~\ref{sec:notation}
and prove our theorem in section~\ref{sec:theorem}.
An alternative to the method mentioned in section~\ref{sec:theorem}
would be to identify \textit{all} the symmetry-constrained NHDM models
for a given $N$; and, within those, all the possible vacua.
One would then
study the existence (or lack thereof) of a decoupling limit
for each case.
This was the method mentioned in \cite{Faro:2020qyp} in connection with the
2HDM.
The 3HDM is the only other case where all symmetry-constrained models
\cite{Ivanov:2012fp} and their respective vacua \cite{Ivanov:2014doa}
have been identified.
We present that alternative (and long) calculation in detail
in section~\ref{sec:3HDM}.
Of course, it confirms our general theorem, but it highlights how simple
and elegant the general result is.
We draw our conclusions in section~\ref{sec:concl}

\section{\label{sec:notation}Notation}

\subsection{\label{sec:potential}The scalar potential}

Consider a  $SU(2)_L \times U(1)_Y$ gauge theory with
$N$ scalar doublets $\Phi_i$ with hypercharge $Y=1/2$
($Q = T_3 + Y$).
The scalar potential can be written as
\cite{Botella:1994cs}
\begin{equation}
V_H =
Y_{ij}\left(\Phi_i^\dagger \Phi_j\right)
+ Z_{ij,kl} \left(\Phi_i^\dagger \Phi_j\right)\left(\Phi_k^\dagger \Phi_l\right),
\label{eq:potential}
\end{equation}
where, by Hermiticity,
\begin{equation}
    Y_{ij}=Y_{ji}^* , \quad
    Z_{ij,kl}=Z_{kl,ij}=Z_{ji,lk}^*.
\end{equation}
We take the vacuum expectation values (vev) which preserve electromagnetism
\begin{equation}
\langle \Phi_i \rangle = 
\begin{pmatrix}
0 \\ v_i/\sqrt{2}
\end{pmatrix}\, ,
\label{eq:vevs}
\end{equation}
which may in general be complex.
The stationary conditions are
\begin{equation}
\left(Y_{ij}+Z_{ij,kl}v_k^* v_l\right)v_j=0.
\label{eq:stationarity}
\end{equation}
Except where indicated otherwise,
we use implicit summation of repeated indices.
Since the mass matrix for the charged scalars is
\begin{equation}
\left(M^2_\pm\right)_{ij} =
Y_{ij}+Z_{ij,kl}v^{*}_k v_l,
\label{eq:mass_matrix_charged}
\end{equation}
the stationarity conditions in \eqref{eq:stationarity}
may be rewritten as
\be
\left(M^2_\pm\right)_{ij} v_j = 0.
\label{eq:null_mass_state}
\ee
This expression will turn out to be quite useful.

\subsection{\label{subsec:basis}Basis transformations and symmetries}

It is very important to make clear the distinct concepts of basis
transformations, on the one hand, and of symmetries, on the other.
We start with the former.
The theory was originally written in terms of fields $\Phi_i$.
These may be traded for
new fields $\Phi^\prime_i$;
\begin{equation}
\Phi_i \rightarrow \Phi_i^{\prime}= U_{ij} \Phi_j.
\label{basis_transformation}
\end{equation}
This is a \textit{basis transformation}
which leaves the kinetic terms unchanged,
making $U$ a $N \times N$ unitary matrix.
Under this transformation,
the potential's parameters and the vevs become
\begin{align}
Y_{ij}&
\rightarrow Y_{ij}^{\prime}= U_{ik}Y_{kl}U_{jl}^*,
\label{eq:newbasis_2} \\
Z_{ij,kl}&
\rightarrow Z_{ij,kl}^{\prime}=U_{im}U_{ko}Z_{mn,op}U_{jn}^*U_{lp}^*,
\label{eq:newbasis_4} \\
v_i& \rightarrow v^{\prime}_i= U_{ij} v_j.
\end{align}
Since such a transformation can have no effect on the physical
predictions,
only basis invariant combinations can be observed experimentally
\cite{Botella:1994cs}.

We now turn to the concept of \textit{symmetry}.
Take
\begin{equation}
\Phi_i \rightarrow \Phi_i^S= S_{ij} \Phi_j,
\label{S_symmetry}
\end{equation}
where $S$ is also a $N \times N$ unitary matrix.
If \eqref{S_symmetry} is indeed a symmetry of the potential
\eqref{eq:potential}, then
\begin{align}
Y_{ij}&
=Y_{ij}^{S}= S_{ik}Y_{kl}S_{jl}^*,
\label{eq:S_on_Y}\\
Z_{ij,kl}&
=Z_{ij,kl}^{S}=S_{im}S_{ko}Z_{mn,op}S_{jn}^*S_{lp}^*.
\label{eq:S_on_Z}
\end{align}
The symmetry may (or not) be spontaneously broken,
depending on whether (or not)
\begin{equation}
v_i = v_i^S= S_{ij} v_j.
\label{eq:Sv}
\end{equation}
The crucial difference between a 
basis transformation and a symmetry is that in the former
the potential parameters do not remain the same,
while in the latter those coefficients must remain
invariant.

Consider a theory in which $V_H$, when written in terms of the fields
$\Phi_i$, has the symmetry $S$.
Now, perform the basis transformation in eq.~\eqref{basis_transformation}.
When written in terms of the new fields $\Phi^\prime_i$,
$V_H$ is no longer invariant under $S$; rather, it is now invariant
under
\begin{equation}
S^\prime=U S U^\dagger.
\label{Sprime}
\end{equation}

\subsection{\label{subsec:CHB}The charged Higgs basis}

The mass matrix for the charged scalars in eq.~\eqref{eq:mass_matrix_charged}
can be diagonalized via a unitary $N \times N$ matrix $U^\textrm{ch}$.
But the basis freedom in eq.~\eqref{basis_transformation}
also involves a unitary $N \times N$ matrix.
Thus,
we may perform a basis change into a basis where the charged components
of each doublet already correspond to mass eigenstates:
\begin{equation}
\Phi^\textrm{ch}_1 =
\left(
\begin{array}{c}
G^+\\*[2mm]
\tfrac{1}{\sqrt{2}}
\left( v + H^0 + i G^0 \right)
\end{array}
\right),
\hspace{5ex}
\Phi^\textrm{ch}_k =
\left(
\begin{array}{c}
H^+_k\\*[2mm]
\tfrac{1}{\sqrt{2}} \varphi^{C0}_k
\end{array}
\right)\, ,
\label{fields_CHB}
\end{equation}
where $H_k^+$ ($ k=2 \ldots N$) are the physical charged Higgs mass eigenstate fields,
with corresponding masses $m^2_{\pm, k}$.
$H_1^\pm=G^\pm$ is the massless would-be Goldstone boson; $m_{\pm, 1}^2=0$.
This is known as the \textit{Charged Higgs basis} (CH basis)
\cite{Bento:2017eti,Nishi:2007nh}.
In this basis,
only the first doublet has a vev,
\begin{equation}
v^\textrm{ch}_1=v\, ,\ \ \ \ \ v^\textrm{ch}_{k\neq 1}=0.
\label{eq:vev_CHB}
\end{equation}
Thus, the CH basis is a particular case of what was dubbed a
``Higgs basis'' in ref.~\cite{Botella:1994cs}.
The matrix which performs the transformation from the original basis
into the CH basis clearly satisfies
\begin{equation}
U^\textrm{ch}_{1 k} = \frac{v_k^\ast}{v}.
\label{eq:Uch_first_line}
\end{equation}

We write the potential in the CH basis as
\begin{equation}
V_H =
Y^\textrm{ch}_{ij}
\left(\Phi_i^{\textrm{ch}\, \dagger} \Phi^\textrm{ch}_j\right)
+ Z^\textrm{ch}_{ij,kl}
\left(\Phi_i^{\textrm{ch}\, \dagger} \Phi^\textrm{ch}_j\right)
\left(\Phi_k^{\textrm{ch}\, \dagger} \Phi^\textrm{ch}_l\right).
\label{eq:potential_CHB}
\end{equation}
The quadratic and quartic coefficients in the CH basis are obtained
by substituting $U$ with $U^\textrm{ch}$ in
eqs.~\eqref{eq:newbasis_2}-\eqref{eq:newbasis_4},
respectively.
The matrix of the charged scalars of eq.~\eqref{eq:mass_matrix_charged}
becomes, in the CH basis,
\begin{eqnarray}
\left(M^{2}_\pm\right)^\textrm{ch}_{ij} &=&
Y^\textrm{ch}_{ij}+Z^\textrm{ch}_{ij,kl}
\left(v^\textrm{ch}_k\right)^{*} v^\textrm{ch}_l
\label{eq:mass_matrix_charged_CHB_line1}
\\
&=&
\delta_{ij}\, m_{\pm, i}^2
\ \ \ \textrm{(no sum)}.
\label{eq:mass_matrix_charged_CHB}
\end{eqnarray}
Recall that we have chosen the transformation $U^\textrm{ch}$ precisely
such that the mass matrix is diagonal and the last equality holds.

As in any basis, the stationarity conditions may still be written as
\be
\left(M^2_\pm\right)^\textrm{ch}_{ij} v^\textrm{ch}_j = 0\, ,
\label{eq:null_mass_state_CHB}
\ee
\textit{c.f.} eq.~\eqref{eq:null_mass_state}.
It is clear from eqs.~\eqref{eq:vev_CHB} and \eqref{eq:mass_matrix_charged_CHB}
that eq.~\eqref{eq:null_mass_state_CHB} indeed holds.

\subsection{\label{subsec:decouple}Decoupling}

As shown in ref.~\cite{Faro:2020qyp},
the CH basis is particularly useful when investigating the decoupling limit.
Looking back at eq.~\eqref{fields_CHB},
if one wishes to decouple the extra doublets, one merely needs to
take the masses $m^2_{\pm,\; k}$ ($k \geq 2$) to be much larger than
$v^2$.
Indeed, it was shown in \cite{Faro:2020qyp} that taking the charged
scalars very massive makes all extra neutral scalars very massive,
and, simultaneously, suppresses any CP violation in scalar-pseudoscalar mixing.

How can one make $(M^2_\pm)^\textrm{ch}$ very large?
Inspecting eq.~\eqref{eq:mass_matrix_charged_CHB_line1},
one might think that there could be various ways to achieve that.
However, this may only be achieved by making 
$Y^\textrm{ch}$ large.\footnote{We are being slightly cavalier
in this definition. Indeed, $m^2_{\pm,\, 1}=0$ cannot be
``made large'', and neither can
$Y_{11}^\textrm{ch}$, nor $Y_{i\ j\neq i}^\textrm{ch}$.
In fact, the latter must obey
$Y_{11}^\textrm{ch} + Z_{11,11}^\textrm{ch} v^2 = 0$
and $Y_{i\ j\neq i}^\textrm{ch} + Z_{ij,11}^\textrm{ch} v^2 = 0$.
But this subtlety does not affect our argument, so
we'll steer clear from overly well defined yet rather
convoluted details.
Whenever we mention ``schematically'' in the text,
this is the detail we have in mind.}
The point is that the quartic coefficients $Z^\textrm{ch}$
are quite constrained by unitarity and perturbativity arguments.
Common estimates take these to lie below $4 \pi$ or $8 \pi$,
with more precise statements possible --
see, for example, ref.~\cite{Bento:2017eti}.
So, the decoupling limit may be written schematically as
\begin{equation}
M_\pm^{2\, \textrm{ch}} =
Y^\textrm{ch}
+
Z^\textrm{ch}
v^{\textrm{ch} \ast}
v^\textrm{ch}
\ \ 
\xrightarrow{\ \textrm{decoupling}\ }
\ \ 
M_\pm^{2\, \textrm{ch}} =
Y^\textrm{ch}\, .
\label{eq:decouple}
\end{equation}
Of course,
the effective decoupling hinges on the possibility that
(again schematically) 
$Y^\textrm{ch}$ can be chosen much larger than $v^2$.
Is this still possible in a symmetry-constrained potential?
This is what we turn to next.

\section{\label{sec:theorem}Theorem and proof}

Let us imagine that the potential in eq.~\eqref{eq:potential} is constrained
by requiring it invariant under a symmetry $S$, as in
eqs.~\eqref{eq:S_on_Y}-\eqref{eq:S_on_Z}.
Then, according to eq.~\eqref{Sprime},
the potential \eqref{eq:potential_CHB} in the CH basis is invariant under
\begin{equation}
S^\textrm{ch} = U^\textrm{ch}\, S\, U^{\textrm{ch}\, \dagger}.
\label{Sch}
\end{equation}
In particular,
\be
Y^\textrm{ch}
=
S^\textrm{ch}\, Y^\textrm{ch}\, S^{\textrm{ch}\, \dagger}.
\label{eq:Sch_on_Ych}
\ee
Eq.~\eqref{Sch} is the crucial observation that has been missed before and,
in particular, in ref.~\cite{Faro:2020qyp}.
We can learn quite a great deal by combining the simplicity of the CH basis
with the form of the symmetry when \textit{written in the CH basis}.

The possibility that the symmetry is not (is) spontaneously broken
depends on whether (or not)
\be
S^\textrm{ch}_{ij} v^\textrm{ch}_j = v^\textrm{ch}_i\, .
\label{eq:Sv_CHB}
\ee
This is just the CH basis version of eq.~\eqref{eq:Sv}.
Since $S^\textrm{ch}$ is $N \times N$ unitary and $v^\textrm{ch}$
satisfies eq.~\eqref{eq:vev_CHB},
one can show that eq.~\eqref{eq:Sv_CHB} holds if and only if $S^\textrm{ch}$
is of the form
\be
S_\textrm{vev\ preserving}^\textrm{ch}
=
 \left( 
    \begin{array}{c|c} 
      
      1 & 0 \\ 
      \hline 
       0 & 
      \begin{matrix}
          &&&\\
          &&\tilde{S}^\textrm{ch}&\\
          &&&\\
      \end{matrix}
    \end{array} 
    \right)\, ,
\label{eq:Sch_restricted}
\ee
where $\tilde{S}^\textrm{ch}$ is any unitary $(N-1) \times (N-1)$ matrix.
Notice that this must hold irrespectively of the specific form of
the symmetry $S$ chosen in the original basis.
All such symmetries of the vacuum will map in the CH basis into symmetries
$S^\textrm{ch}$ of the form \eqref{eq:Sch_restricted}.
Conversely,
all symmetries of the type \eqref{eq:Sch_restricted},
will through $S = U^{\textrm{ch}\, \dagger}\, S^\textrm{ch}\, U^\textrm{ch}$
map in the original basis onto symmetries of the vacuum,
where eq.~\eqref{eq:Sv} holds.

It is important to take a brief detour here.
The last paragraph means that there is a very large set of symmetries of the
vacuum in the original basis; a set that can be mapped onto $SU(N-1)$.
Imagine that one has a group of symmetries $\{S_1, S_2, \dots \}$ of the Lagrangian,
and one wishes to inquire whether they survive spontaneous symmetry breaking (SSB).
If this group is small, then it is conceivable that a vacuum may be found
that breaks them all.
In contrast,
if the Lagrangian is invariant under a very large group,
and given the fact that there are so many possible invariances of the vacuum,
it becomes unlikely or even impossible for all symmetries of the Lagrangian
to be spontaneously broken.
Said otherwise, 
for a large enough group there will always be some remnant symmetries after SSB.
This is mentioned explicitly for the 3HDM in section 5.2 of
ref.~\cite{Ivanov:2014doa}.
Our eq.~\eqref{eq:Sch_restricted} shows why this must be true in general.
Moreover,
the existence or not of remnant symmetries is very important for a full theory
including fermions.
Indeed, it was shown in refs.~\cite{Leurer:1992wg} and \cite{Felipe:2014zka} that
the only way of obtaining a physical CKM mixing matrix and,
simultaneously, non-degenerate and non-zero quark masses is to
require that the vevs of the Higgs fields break
completely the full flavour group.
This ends our detour into the importance of eq.~\eqref{eq:Sch_restricted}.

We now state our:\\
\textbf{Theorem:} Given a Lagrangian with symmetry $S$,
the theory has a decoupling limit if and only if the vacuum
also has the same symmetry $S$.

Let us start our proof by assuming that $S$ is a symmetry of the Lagrangian.
Then,
in the CH basis,
eq.~\eqref{eq:Sch_on_Ych} holds.
We now assume that there is a decoupling limit,
in the form of eq.~\eqref{eq:decouple}.
Combining,
we find
\be
(M_\pm^2)^\textrm{ch}
=
S^\textrm{ch}\, (M_\pm^2)^\textrm{ch}\, S^{\textrm{ch}\, \dagger}.
\label{eq:Sch_on_Mch_decouple}
\ee
%
But eq.~\eqref{eq:Sch_on_Mch_decouple} holds
if and only if $S^\textrm{ch}$ is of the form
\eqref{eq:Sch_restricted}.
(We will even derive much stronger implications after eq.~\eqref{eq:crucial_1} below.)
This, in turn, holds if and only if $S^\textrm{ch}$ is a symmetry
of the vacuum $v^\textrm{ch}$, as mentioned above.
Thus, $S$ is a symmetry of
$v_k$ in the original basis.
In short, if $S$ is a symmetry of the Lagrangian and we require a decoupling limit,
then $S$ must also be a symmetry of the vacuum.

Conversely,
imagine that $S$ is a symmetry of the Lagrangian and a symmetry of the vacuum.
Then, we know from eq.~\eqref{eq:mass_matrix_charged} that $S$ is also
a symmetry of the charged mass matrix.
Thus,
eq.~\eqref{eq:Sch_on_Mch_decouple} holds in the CH basis,
with $S^\textrm{ch}$ given in eq.~\eqref{Sch}.
The question now is whether 
eq.~\eqref{eq:Sch_on_Mch_decouple} allows or not for decoupling.
We start by writing eq.~\eqref{eq:Sch_on_Mch_decouple} as
the commutator equation
\be
\left[
(M_\pm^2)^\textrm{ch} ,\ 
S^\textrm{ch}
\right] = 0\, .
\label{eq:crucial_1}
\ee
Given eq.~\eqref{eq:mass_matrix_charged_CHB},
this translates into
\be
\left( m_{\pm, i}^2 - m_{\pm, j}^2\right)\, 
S^\textrm{ch}_{ij} = 0
\ \ \ \textrm{(no sum)}\, .
\label{eq:crucial_2}
\ee
If $i=1$ and $j \neq 1$,
then, because $G^\pm$ has mass $m_{\pm, 1}^2=0$,
while $m_{\pm, j}^2 \neq 0$ (assume this for the moment),
one concludes that $S^\textrm{ch}_{1j} = 0$.
Similarly, $S^\textrm{ch}_{i1} = 0$.
Moreover,
since $S^\textrm{ch}$ is a unitary matrix,
one is forced into $S^\textrm{ch}_{11}=1$,
and $S^\textrm{ch}$ must have the form in eq.~\eqref{eq:Sch_restricted}.
This is the assertion we have used in the previous paragraph.

Eq.~\eqref{eq:crucial_2} also means that if $S^\textrm{ch}$
has any nonzero entry with $i \neq j$, then
the corresponding charged scalars \textit{must be degenerate}.
In particular, one might consider theories with symmetries
$S$ corresponding to $S^\textrm{ch}_{1, j\neq 1} \neq 0$.
But those theories would have more than one massless scalar field,
and, thus, be ruled out by experiment.\footnote{The presence of
a second massless charged scalar field could possibly
be solved by including an extra gauge group, of which this would be
the corresponding would-be Goldstone boson. We will not consider
that possibility here.}
Excluding those cases,
the charged scalars have masses which may be taken to infinity in a way
consistent with the symmetry $S^\textrm{ch}$ in the CH basis or
(which is the same) $S$ in the original basis.
Thus, the theory does have a decoupling limit.
In short, if $S$ is a symmetry of the Lagrangian and of the vacuum,
then the theory has a decoupling limit.

Given our theorem,
and starting from a scalar potential invariant under $S$,
eq.~\eqref{eq:crucial_1} can be viewed as an effective definition of decoupling.
Indeed, since commuting matrices do so in any basis, an alternative
definition would be
\be
\left[
M_\pm^2 ,\ S
\right] = 0\, .
\label{eq:crucial_3}
\ee
Of course, it is much simpler to check whether or not $S v = v$,
as proposed in the theorem. Still eq.\eqref{eq:crucial_3} is interesting.

Eq.~\eqref{eq:crucial_2} is even more powerful than it seems.
It tells us exactly how the decoupling might be achieved.
Let us concentrate on $i \neq 1$ and $j \neq 1$.
To do so we, think of $\tilde{S}^\textrm{ch}_{ab}$,
where $a,b = 2, 3, \dots N$,
thus sidestepping the Goldstone boson issues already discussed.
For a given $a \neq b$ there are two possibilities
\begin{enumerate}
\item
$S$ is such that $\tilde{S}^\textrm{ch}_{ab} \neq 0$
\ \ \ $\Longrightarrow$\ \ \
$m_{\pm, a}^2 = m_{\pm, b}^2$
are degenerate;
\item
$m_{\pm, a}^2 \neq m_{\pm, b}^2$
are not degenerate
\ \ \ $\Longrightarrow$\ \ \
$S$ is such that $\tilde{S}^\textrm{ch}_{ab}=0$
\end{enumerate}

As an illustration, let us consider the 3HDM to be fully analyzed below.
Requiring decoupling,
there are two possibilities for $S$.
It may lead into
\be
S^\textrm{ch}
=
 \left( 
    \begin{array}{ccc} 
      
      1 & 0 & 0\\ 
      0 & e^{i \alpha}&0\\
      0 & 0 & e^{i \beta}
    \end{array} 
    \right)\, ,
\label{Sch_1}
\ee
in which case the two charged scalars may have different masses
$m_{\pm, 2}^2 \neq m_{\pm, 3}^2$ that can be taken to infinity independently.
This includes in particular the possibility that one decouples the 3HDM into
an effective 2HDM by taking only $m_{\pm, 3}^2 \rightarrow \infty$.
Alternatively,
$S$ may be such that
\be
S^\textrm{ch}
=
 \left( 
    \begin{array}{c|c}   
      1 & 0 \\ 
      \hline 
       0 & 
      \begin{matrix}
          &&&\\
          &&\tilde{S}^\textrm{ch}&\\
          &&&\\
      \end{matrix}
    \end{array} 
    \right)\, ,
\label{Sch_2}
\ee
where now $\tilde{S}^\textrm{ch}$ is a non-diagonal unitary $2 \times 2$
matrix.
Then, $m_{\pm, 2}^2 = m_{\pm, 3}^2$ and the 3HDM can only decouple directly
into the SM,
by taking $m_{\pm, 2}^2 =m_{\pm, 3}^2 \rightarrow \infty$.

This study has an impact on non-abelian symmetry groups.
Indeed, take some symmetry $S_k$ ($k=1, 2, \dots$),
and the corresponding $S^\prime_k$,
obtained from $S_k$ through eq.~\eqref{Sprime}
with the same basis transformation $U$ for all.
If $[S_1, S_2] = 0$ in one basis, then $[S^\prime_1, S^\prime_2] = 0$
in another.
As a result, if the generators of a group are not simultaneously diagonal
in some basis, neither will they be in any other basis.
In particular
\be
[S_1, S_2] \neq 0
\ \ \Longrightarrow\ \ 
[S_1^\textrm{ch}, S_2^\textrm{ch}] \neq 0\, .
\ee
Thus, in such a case, one of the two $S_k^\textrm{ch}$ will have
off-diagonal entries and the decoupling charged scalars
corresponding to those entries will be degenerate.
We will illustrate both cases
\eqref{Sch_1} and \eqref{Sch_2}
with the full analysis of the 3HDM in the next section.

In contrast, abelian groups always permit non-degenerate charged scalar masses.
Indeed, if $[M_\pm^2 , S] = 0$, $S$ is in its diagonal basis,
and $S$ has non-degenerate eigenvalues,
then $M_\pm^2$ is diagonal and we are already in the charged Higgs basis.
If $S$ has a degenerate subspace, then $M_\pm^2$ may be off-diagonal in that subspace.
But bringing $M_\pm^2$ into its CH basis diagonal form will not affect 
the diagonal form of $S$. 
Indeed, in that subspace $S$ is proportional to the unit matrix
and is unaffected by the mass diagonalization eventually required in that subspace.
We are again in the CH basis.
Thus proving our assertion.

\section{\label{sec:3HDM}A full study of 3HDM}

We have already proved for any NHDM that,
given a Lagrangian with symmetry $S$,
the theory has a decoupling limit if and only if the vacuum
also has the same symmetry $S$.
However,
the 3HDM is the only NHDM besides the 2HDM where all the symmetries and
corresponding vacua are known \cite{Ivanov:2012fp,Ivanov:2014doa}.
Thus, it is interesting to redo the proof of our theorem for $N=3$ by:
i) analyzing all possible symmetry-vacua pairs one by one;
ii) studying
their mass matrices (charged and neutral);
and, iii) probing whether they allow (or not)
for decoupling in accordance with the theorem (as they must).
This is also interesting because it will allow us to illustrate some
of the remarks on the exact features of the alignment which we have made
at the end of the previous section.

\subsection{\label{subsec:3HDM_method}General method}

Here, we describe the method used to test whether the masses of the particles
predicted by several 3HDM have a decoupling limit or not.

The inputs to this method are a potential $V_H$ and a respective vev.
\begin{enumerate}
\item
The stationarity equations impose conditions on the parameters of
the potential, whose number depends on the degrees of freedom
the vev has. These conditions will be referred to as $t_a$.
\item
Every doublet is expanded around the vev - \textit{c.f.} eq.~\eqref{eq:exp_vev}
- and substituted back in the potential, such that the potential
will have extra functional dependencies - \textit{c.f.} eq.~\eqref{eq:func_depend}:
\begin{equation}
    \Phi_i = 
    \begin{bmatrix}
    \varphi^+_i \\
    v_i + (H_i+\chi_i)/ \sqrt{2}
    \end{bmatrix}\, ,
    \label{eq:exp_vev}
\end{equation}
\begin{equation}
    V_H = V_H(\varphi^+_i,\varphi^-_i, v_i, H_i, \chi_i)\, .
    \label{eq:func_depend}
\end{equation}
\item
The mass (squared) matrices are calculated as being the
Hessian of the potential in two different subspaces;
the charged subspace and the neutral one:
\begin{equation}
    (M^2_\pm)_{ij} =
    \frac{\partial^2V_H}{\partial \varphi^+_i
    \partial \varphi^-_j }\bigg|_{\{t_a\},
    (\varphi^+_b   \varphi^-_b, H_b, \chi_b)\rightarrow 0}\, ,
\end{equation}
\begin{equation}
   (M^2_\textrm{neutral})_{ij} =
    \frac{\partial^2V_H}{\partial (H,\chi)_i \partial (H,\chi)_j }
    \bigg|_{\{t_a\},(\varphi^+_b   \varphi^-_b, H_b, \chi_b)
    \rightarrow 0}\, ,
\end{equation}
where, recall, $t_a$ are the conditions obtained from minimization
of the potential in step 1.
\item
The eigenvalues of these matrices are the (squared) masses of the
charged scalars in the first subspace, ($\varphi^+, \varphi^-$),
and of the neutral scalars in the second, ($H,\chi$).
\item
Whether the masses have a decoupling limit or not can only be decided
by looking at the parametrical dependence of the eigenvalues.
If any mass depends on a free parameter such as $m_i$ (where $i= 1,2,3$),
then there is a decoupling limit.
Otherwise, the masses are said to have non-decoupling.
\end{enumerate}

It is often the case where the matrix in the subspace ($H,\chi$)
isn't diagonalizable analytically.
Indeed,
taking the obvious massless Goldstone boson out of the matrix,
this still involves the solution of a polynomial of degree five.
In such cases
we evaluate the decoupling limit using the trace of the matrix.
This works because the trace of the matrix is the sum of its eigenvalues,
all of which are masses squared and, thus positive.
Thus, the trace can be taken to infinity if and only if there is at least
one massive state which can.
In contrast,
if no mass can decouple, then neither will the trace.

\subsection{\label{subsec:3HDM_examples}Some simple examples}

Using the method described above,
we can now show more concretely what is decoupling and non-decoupling.
To this end we apply the method to a three Higgs doublet model
with a ${\mathbb{Z}}_2 \rtimes {\mathbb{Z}}_2 \rtimes \mathbb{Z}^*_2$ symmetry.
The potential for the model may be written as
\begin{eqnarray}
V_H &=&
-\sum_{1 \leq i \leq 3} m_{i}^{2}\left(\phi_{i}^{\dagger}
\phi_{i}\right)+\sum_{1 \leq i \leq j \leq 3}
\lambda_{i j}\left(\phi_{i}^{\dagger} \phi_{i}\right)
\left(\phi_{j}^{\dagger} \phi_{j}\right)+\sum_{1 \leq i<j \leq 3}
\lambda_{i j}^{\prime}\left(\phi_{i}^{\dagger}
\phi_{j}\right)\left(\phi_{j}^{\dagger} \phi_{i}\right)
\nonumber\\ 
&&
+ \lambda_{1}\left(\phi_{2}^{\dagger} \phi_{3}\right)^{2}
+\lambda_{2}\left(\phi_{3}^{\dagger} \phi_{1}\right)^{2}
+\lambda_{3}\left(\phi_{1}^{\dagger} \phi_{2}\right)^{2} + H.c.
\end{eqnarray}
All possible vev's for this model (and to all other
realizable symmetry-constrained 3HDM models)
can be found in \cite{Ivanov:2014doa}.

First, as an example of decoupling, we take a vev that does not break the
${\mathbb{Z}}_2 \rtimes {\mathbb{Z}}_2 \rtimes \mathbb{Z}^*_2$ symmetry;
the vev $(v, 0, 0)$ (with $v$ real).
Following the method described in section~\ref{subsec:3HDM_method},
we obtain the mass matrices.
For the charged fields ($\varphi^{\pm}_i$), the mass matrix is
\begin{equation}
    \left(
\begin{array}{ccc}
 0 & 0 & 0 \\
 0 & \frac{\lambda_{12} {v}^2}{2}-m_2 & 0 \\
 0 & 0 & \frac{\lambda_{13} {v}^2}{2}-m_3 \\
\end{array}
\right)\, .
\label{eq:decoupling_model_1}
\end{equation}
The mass matrix for the neutral fields (i.e. $H_i$ and $\chi_i$) is
\begin{equation}
 \left(
\begin{array}{cccccc}
 2\lambda_{11} {v}^2 & 0 & 0 & 0 & 0 & 0 \\
 0 & \frac{{v}^2}{2}(\lambda_{12} +\lambda'_{12})-m_2 & 0 & 0 & 0 & 0 \\
 0 & 0 & \frac{{v}^2}{2}(\lambda_{13} +\lambda'_{13})-m_3 & 0 & 0 & 0 \\
 0 & 0 & 0 & 0 & 0 & 0 \\
 0 & 0 & 0 & 0 & \frac{{v}^2}{2}(\lambda_{12} +\lambda'_{12})-m_2 & 0 \\
 0 & 0 & 0 & 0 & 0 & \frac{{v}^2}{2}(\lambda_{13} +\lambda'_{13})-m_3 \\
\end{array}
\right)  \, . 
\end{equation}
The matrices obey the minimum condition $m_1=\lambda_{11} {v}^2$.
We see that both matrices are immediately diagonal. And, from the eigenvalues,
we notice that the fields $\Phi_2$ and $\Phi_3$ can decouple,
because both have a free parameter
($m_2$ and $m_3$, respectively) that can be taken to be arbitrarily large.
In this case, the vev does not break the symmetry and there is decoupling.

For the second example, we take the vev $(0, v_2, v_3)$ for the case $\lambda_1 < 0$.
We note that this vev breaks the
${\mathbb{Z}}_2 \rtimes {\mathbb{Z}}_2 \rtimes \mathbb{Z}^*_2$ symmetry,
leaving a residual symmetry of the type ${\mathbb{Z}}_2 \rtimes \mathbb{Z}^*_2$.
Following the method described in section~\ref{subsec:3HDM_method},
we obtain non-diagonal matrices.
The mass eigenvalues for the charged fields are
\begin{equation}
    \left\{0,-\frac{1}{2} (2 \lambda_1+\lambda'_{23}) \left({v_2}^2+{v_3}^2\right),\frac{1}{2} \left({\lambda_{12}} {v_2}^2+{\lambda_{13}} {v_3}^2-2 {m_1}\right)\right\}\, ,
\end{equation}
with corresponding eigenvectors
\begin{equation}
 \left\{(0, \frac{v_2}{v_3}, 1), (0, -\frac{v_3}{v_2}, 1), (1, 0, 0) \right\}\, ,
\end{equation}
in the basis ($\varphi^{+}_1$, $\varphi^{+}_2$, $\varphi^{+}_3$).
Note that the first two eigenvectors need a normalization constant,
which is irrelevant for our purposes.

For the neutral fields the eigenvalues are
\begin{eqnarray}
&&
\Big\{ 0, \quad -2 \left({\lambda_1} {v_2}^2+{\lambda_1} {v_3}^2\right),
\frac{1}{2} \left({v_2}^2 ({\lambda_{12}}+
{\lambda'_{12}}-2 {\lambda_3})+{v_3}^2 ({\lambda_{13}}+
{\lambda'_{13}}-2 {\lambda_2})-2 {m_1}\right),
\nonumber\\
&&
\hspace{2mm}
\frac{1}{2} \left({v_2}^2 ({\lambda_{12}}+{\lambda'_{12}}
+2 {\lambda_3})+{v_3}^2 ({\lambda_{13}}+{\lambda'_{13}}
+2 {\lambda_2})-2 {m_1}\right),
\nonumber\\ 
&&
\hspace{2mm}
- \sqrt{{v_ 2}^2 {v_ 3}^2 (2 {\lambda_1}+{\lambda_{23}}+{\lambda'_{23}})^2
+\left({\lambda_{22}} {v_ 2}^2-{\lambda_{33}}
{v_ 3}^2\right)^2}+{\lambda_{22}} {v_ 2}^2
+{\lambda_{33}} {v_ 3}^2, 
\nonumber\\ 
&&
\hspace{2mm}
\sqrt{{v_ 2}^2 {v_ 3}^2 (2 {\lambda_1}+{\lambda_{23}}
+{\lambda'_{23}})^2+\left({\lambda_{22}} {v_ 2}^2
-{\lambda_{33}} {v_ 3}^2\right)^2}
+{\lambda_{22}} {v_ 2}^2+{\lambda_{33}} {v_ 3}^2
\Big\}\, .
\end{eqnarray}
The eigenvectors associated with these eigenvalues are,
in the basis $(H_1, H_2, H_3, \chi_1, \chi_2, \chi_3)$,
\begin{eqnarray}
&& \Bigg\{\left\{0,0,0,0,\frac{{v_2}}{{v_3}},1\right\},
\left\{0,0,0,0,-\frac{{v_3}}{{v_2}},1\right\},\{0,0,0,1,0,0\},\{1,0,0,0,0,0\},
\nonumber\\ 
&&
\hspace{2mm}
\left\{  0,\frac{-\sqrt{{v_ 2}^2 {v_ 3}^2 (2 {\lambda_1}
+{\lambda_{23}}+{\lambda'_{23}})^2
+\left({\lambda_{22}} {v_ 2}^2-{\lambda_{33}} {v_ 3}^2\right)^2}
+{\lambda_{22}} {v_ 2}^2-{\lambda_{33}} {v_ 3}^2}{{v_ 2} {v_ 3} 
(2 {\lambda_1}+{\lambda_{23}}+{\lambda'_ {23}})},1,0,0,0\right\},
\nonumber\\
&&
\hspace{2mm}
\left\{0,\frac{\sqrt{{v_ 2}^2 {v_ 3}^2
(2 {\lambda_1}+{\lambda_{23}}+{\lambda'_{23}})^2
+\left({\lambda_{22}} {v_ 2}^2-{\lambda_{33}} {v_ 3}^2\right)^2}
+{\lambda_{22}} {v_ 2}^2-{\lambda_{33}} {v_ 3}^2}{{v_ 2} {v_ 3}
(2 {\lambda_1}+{\lambda_{23}}+{\lambda'_ {23}})},1,0,0,0\right\}\Bigg\}\, .
\end{eqnarray}
We see from the eigenvalues that two of the fields do not decouple
(these are a mixture of $\Phi_2$ and $\Phi_3$ in the eigenbasis),
since there is no $\lambda$ free term that we can make arbitrarily large.
Here we see that the vev breaks the symmetry of the model
and there is no decoupling limit.

We have only shown explicitly two simple examples.
There are other possible vev's in the
${\mathbb{Z}}_2 \rtimes {\mathbb{Z}}_2 \rtimes \mathbb{Z}^*_2$ model.
However,
some of them will result in very complicated mass matrices,
where determining analytically the eigenvalues is no longer possible.
In such cases, it is necessary to evaluate the trace of
the matrices to see if there is decoupling or not,
as described at the end of the previous section.

\subsection{\label{subsec:3HDM_table}Exhaustive list of symmetry-constrained 3HDM}

Following the method and examples above,
we have studied all the symmetry-vacua pairs identified in
ref.~\cite{Ivanov:2014doa}.
Our results are shown in Table~\ref{table:3HDM_full}.
%
\begin{table}[h]
\begin{tabular}{l|l|l|l}
Group G & vev & Breaks G?                                                                                                                                                                   & Decoupling ? \\ \hline
${\mathbb{Z}}_2 \rtimes {\mathbb{Z}}_2 \rtimes \mathbb{Z}^*_2$  & ($v$,0,0)                                                                                                         &  No     & Y\\ 
& (0, $v_2 e^{i\frac{\pi}{4}}$, $v_3 e^{-i\frac{\pi}{4}})$  & Y     & No\\
& (0, $v_2$, $ \pm v_3)$ & Y     & No\\
& ($v_1 e^{i\frac{k_1 \pi}{2}}$, $v_2e^{i\frac{k_2 \pi}{2}}$,
$v_3 e^{i\frac{k_3 \pi}{2}})$, $k_i \in \mathbb{Z}$& Y     & No \\ 
& ($v_1 e^{i\xi_1}$, $v_2e^{i\xi_2}$, $v_3 e^{i\xi_3})$,
$\xi_i = \xi_i(v_1,v_2,v_3,\lambda)$   & Y & No
\\ \hline
${\mathbb{Z}}_3 \rtimes \mathbb{Z}^*_2$ & ($v$,0,0) & No & Y\\
& ($v_1$,$v_2$,0) & Y     & No\\ 
& ($v_1$, $v_2 e^{i\frac{k_2 \pi}{3}}$,
$v_3e^{i\frac{k_3 \pi}{3}}$), $k_i \in \mathbb{Z}$  & Y     & No\\ 
& ($v_1 e^{i\xi_1}$, $v_2e^{i\xi_2}$, $v_3 e^{i\xi_3})$,
$\xi_i = \xi_i(v_1,v_2,v_3,\lambda)$ & Y     & No
\\ \hline
${\mathbb{Z}}_4 \rtimes \mathbb{Z}^*_2$  & ($v$,0,0) &   No  & Y\\
& (0, $v_2 e^{i\frac{k_2\pi}{4}}$,$v_3 e^{i\frac{k_3\pi}{4}}$),
$k_i \in \mathbb{Z}$ & Y & No\\
& ($v_1$, $\pm v_2 e^{\mp i\frac{k\pi}{4}}$,$\mp v_3 e^{\mp i\frac{k\pi}{4}}$),
$k \in \mathbb{Z}$ & Y & No
\\ \hline
$D_4$    & ($v$,0,0)&   No    & Y\\ 
& $(v_1,v_2,v_3)$ & Y     & No\\
& $(v_1,\pm v_2 e^{i \xi}, \pm v_2 e^{- i \xi})$ & Y     & No\\
& $(v_1,v_2,iv_3)$ & Y     & No
\\ \hline
$S_3$ & $(v,0,0)$&   No    & Y\\
& $(v_1,v_2,v_3) $ & Y     & No\\
& $(v_1,v_2 e^{i \xi},v_2e^{i \xi})$ & Y     & No
\\ \hline
$S_4$    & $(v,0,0)  $ & Y     & No\\
& $(v,v,v) $ & Y     & No\\
& $(\pm v, v \omega, v \omega^2)$ & Y     & No\\
& $(0,v,i v) $ & Y     & No
\\ \hline
$A_4$ & $(v,0,0)$ & Y  & No\\
& $(v,v,v)$ & Y     & No\\
& $(\pm v, v\omega, v \omega^2)$ & Y     & No\\
& $(0,v,v e^{i \alpha})$& Y     & No
\\ \hline
$\Delta(27)$ \textbf{family}  &  $(v\omega,v,v)$& Y  & No\\
& $(v\omega^2,v,v)$ & Y  & No\\
& $(v,0,0)$ & Y   & No\\
& $(v,v,v)$ & Y     & No
\end{tabular}
\caption{\label{table:3HDM_full}List
of all symmetry-constrained models via Higgs family symmetries
in the 3HDM with corresponding vacua, from \cite{Ivanov:2012fp,Ivanov:2014doa}.
The third column indicates whether or not the vacuum breaks the symmetry.
For each pair, we have found the charged and neutral scalar mass matrices 
and (as explained in the text) have identified whether or
not there is decoupling. This is noted in the fourth column.}
\end{table}

There are several things to note in Table~\ref{table:3HDM_full}.
The parameter $\omega = e^{i\frac{2 \pi}{3}}$, meaning that $\omega^3=1$.
The parameter $\lambda$ in $\xi_i = \xi_i(v_1,v_2,v_3,\lambda)$ stands
for all the coupling parameters in the potential.
The dependence of $\xi_i$ on these parameters can be determined
following the procedure described in \cite{Ivanov:2014doa}.
There are vevs that were not written down since they reduce
trivially to the other cases studied.
There are several other vevs that can be obtained from the ones
listed by the group action; such a collection of vevs is dubbed
orbits in ref.~\cite{Ivanov:2014doa}.
Vevs on the same orbit lead to identical physical consequences.

Finally, from the table we can check that indeed for the 3HDM,
whenever the vev breaks the group symmetry there is no decoupling.
Conversely, if the vev is invariant under the group action,
then there is a decoupling limit.
This is a confirmation of our theorem via an explicit independent method,
albeit only possible for the 3HDM.

\subsection{\label{subsec:3HDM_decoupling_cases}The symmetry-constrained 3HDM
models with decoupling}

Inspection of Table~\ref{table:3HDM_full} shows that there are only
five symmetry-vacua pairs which do allow for a decoupling limit.
These are
\begin{enumerate}
\item ${\mathbb{Z}}_2 \rtimes {\mathbb{Z}}_2 \rtimes \mathbb{Z}^*_2$ with vev
$(v,0,0)$.
The charged scalar masses has been presented in eq.~\eqref{eq:decoupling_model_1}.
This is an abelian group,
and there are two different charged scalar masses which,
in agreement with the discussion after eq.~\eqref{Sch_1},
may be taken to
infinity independently.
\item ${\mathbb{Z}}_3 \rtimes \mathbb{Z}^*_2$ with vev $(v,0,0)$.
The charged scalar mass matrix is
\begin{equation}
 \left(
    \begin{array}{ccc}
       0  & 0 & 0 \\
       0  & -m_2 + v^2 \lambda_{12} & 0 \\
       0  & 0  & -m_3 + v^2 \lambda_{13}
    \end{array}
    \right)\, .
    \label{eq:decoupling_model_2}
\end{equation}
Again, in accordance with the discussion after eq.~\eqref{Sch_1},
there are two different charged scalar masses which
may be taken to infinity independently.
\item ${\mathbb{Z}}_4 \rtimes \mathbb{Z}^*_2$ and vev $(v,0,0)$.
The charged scalar mass matrix is
\begin{equation}
 \left(
    \begin{array}{ccc}
       0  & 0 & 0 \\
       0  & -m_2 + v^2 \lambda_{12} & 0 \\
       0  & 0  & -m_3 + v^2 \lambda_{13}
    \end{array}
    \right)\, ,
\label{eq:decoupling_model_3}
\end{equation}
following the decoupling pattern of the previous two cases.
\item
$D_4$ with vev $(v,0,0)$.
Here the charged scalar mass matrix is
\begin{equation}
 \left(
\begin{array}{ccc}
 0 & 0 & 0 \\
 0 & -m_2^2+\lambda_3 v^2+\lambda_3' v^2 & 0 \\
 0 & 0 & -m_2^2+\lambda_3 v^2+\lambda_3' v^2 \\
\end{array}
\right)\, ,
\label{eq:decoupling_model_4}
\end{equation}
which is degenerate.
Thus, both charged scalar masses must be taken to infinity simultaneously,
and one can only reach the full 3HDM $\rightarrow$ SM decoupling limit.
This is a confirmation of the discussion following eq.~\eqref{Sch_2},
and it is related with the fact that the $D_4$ generators
can be taken as
\be
a_3 = \textrm{diag}(1, i, -i)\, ,
\hspace{6ex}
g_1 =
    \begin{pmatrix}
    1 &0 &0\\
    0 &0 &1\\
    0 &1 &0\\
    \end{pmatrix}\, ,
\ee
which do not commute.
\item
$S_3$ with vev $(v,0,0)$.
Here the charged scalar mass matrix
%
coincides with eq.~\eqref{eq:decoupling_model_4}.
Thus,
as for $D_4$,
the charged scalar masses are degenerate and
the generators of $S_3$,
\be
a_3 = \textrm{diag}(1, \omega, \omega^2)\, ,
\hspace{6ex}
g_1 =
    \begin{pmatrix}
    1 &0 &0\\
    0 &0 &1\\
    0 &1 &0\\
    \end{pmatrix}\, ,
\ee
are also non-commuting.
\end{enumerate}

One might be surprised by the fact that all the vevs in Table~\ref{table:3HDM_full}
that lead to decoupling are $(v,0,0)$ which, by definition,
is equivalent to stating that the fields are already in a Higgs basis.
In fact, noticing that the matrices in 
eqs.~\eqref{eq:decoupling_model_1} and
\eqref{eq:decoupling_model_2}-\eqref{eq:decoupling_model_4},
are already diagonal,
we know that the fields are actually written from the start in the CH basis.
(Recall that the CH basis is a particular case of a Higgs basis.)
As far as we can tell,
this has no profound physical justification.
We know for certain that the vevs could have been written in any other form
in the same orbit, had we changed the form of the symmetry generators.
This is easily illustrated in the 2HDM.
One can study the ${\mathbb{Z}}_2$ group generated by
\be
\left(
\begin{array}{cc}
1 & 0\\
0 & -1
\end{array}
\right)
\ \ \ 
\textrm{with vev}
\ \ \ (v,0),
\label{eq:2HDM_Z2}
\ee
which does lead to decoupling,
or one can study the group generated by\footnote{The model based
on the group generated by the matrix in eq.~\eqref{eq:2HDM_Pi2} 
was dubbed $\Pi_2$ in ref.~\cite{Ferreira:2009wh,HS}.
Of course, it is just ${\mathbb{Z}}_2$ in a different basis.
But the distiction is interesting if one were going to impose
a symmetry under both eq.~\eqref{eq:2HDM_Z2} and eq.~\eqref{eq:2HDM_Pi2}
in the \textit{same basis}.}
\be
\left(
\begin{array}{cc}
0 & 1\\
1 & 0
\end{array}
\right)
\ \ \ 
\textrm{with vev}
\ \ \ (v,v)/\sqrt{2},
\label{eq:2HDM_Pi2}
\ee
which also leads to decoupling.
In fact, the models and vacua in eqs.~\eqref{eq:2HDM_Z2} and \eqref{eq:2HDM_Pi2}
are exactly the same but written in different bases.
Since the vev in eq.~\eqref{eq:2HDM_Pi2} does not correspond to a Higgs basis,
we see that the fact that all the situations in Table~\ref{table:3HDM_full}
which lead to decoupling are already in the CH basis is a red herring.
But, at least in principle, it could happen that all vacua leading
into decoupling should be in the same orbit as a vev with only one nonzero
entry. We see no reason for that, but we cannot exclude it forthright,
so this is an open problem.

\section{\label{sec:concl}Conclusions}

We have studied the decoupling properties of the most general NHDM model.
We have shown a very powerful theorem,
stating that, given a Lagrangian with scalar family symmetry $S$,
the theory has a decoupling limit if and only if the vacuum
also has the same symmetry $S$.
We have also produced an independent proof for the special case of the
3HDM. This was possible because, in the 3HDM, all the symmetry-constrained
realizable models and their vacua are known \cite{Ivanov:2014doa}.
This special 3HDM proof complements a proof along the same lines
for the 2HDM, mentioned in \cite{Faro:2020qyp}.
Producing results along these lines for any NHDM
with $N \geq 4$ would require the knowledge of
all the symmetry-constrained models and corresponding vacua
for those cases as well. This is unknown at the moment and
is certainly exceedingly challenging.
This highlights how elegant our proof for the general NHDM
really is.

Along the way, we proved an interesting result concerning the behavior
of the charged scalar mass matrix $M_\pm^2$ under an exact symmetry.
Symmetry under an abelian group can accommodate non-degenerate
charged scalar masses, while a non-abelian group will \textit{per force} imply
some degeneracy in charged scalar masses.

\begin{acknowledgments}
We are grateful to the anonymous referee of ref.~\cite{Faro:2020qyp} for
asking the question that prompted this work.
This work is supported in part by the Portuguese
Funda\c{c}\~{a}o para a Ci\^{e}ncia
e Tecnologia\/ (FCT) under Contracts
CERN/FIS-PAR/0008/2019,
PTDC/FIS-PAR/29436/2017,
UIDB/00777/2020,
and UIDP/00777/2020;
these projects are partially funded through POCTI (FEDER),
COMPETE,
QREN,
and the EU.
\end{acknowledgments}

%


\end{document}